\documentclass[doublecol]{epl2} 
\usepackage{epsfig}
\usepackage{amsmath}
\usepackage{amssymb}
\title{Nanomechanical Mass Measurement using Nonlinear Response of a Graphene Membrane}
\shorttitle{Nanomechanical Mass Measurement} %Insert here a short version of the title if it exceeds 70 characters

\author{J. Atalaya\inst{1}  \and J. M. Kinaret\inst{1} \and A. Isacsson\inst{1}}
\shortauthor{J. Atalaya \etal}

\institute{                    
  \inst{1} Department of Applied Physics, Chalmers University of Technology, SE-412 96 G{\"o}teborg Sweden.\\
 
}
\pacs{85.85.+j}{Nanoelectromechanical systems}
\pacs{06.30.Dr}{Mass and Density}
\pacs{05.45.-a}{Nonlinear dynamics and chaos}

\abstract{We propose a scheme to measure the mass of a single particle
  using the nonlinear response of a 2D nanoresonator with degenerate
  eigenmodes. Using numerical and analytical calculations, we show
  that by driving a square graphene nanoresonator into the nonlinear
  regime, simultaneous determination of the mass and position of an
  added particle is possible. Moreover, this scheme only requires
  measurements in a narrow frequency band near the fundamental
  resonance.}

\begin{document}

\maketitle

\section{Introduction}
Nanoelectromechanical (NEM) resonators hold promise as ultrasensitive
mass detectors~\cite{Roukes_Review_2005,Boisen_2009}. NEM mass sensors
(NEM-MS) rely on a resonant frequency shift $\Delta\omega$ due to an
added mass $\Delta M$. However, as opposed to detecting a single
adsorbed particle, to actually measure its mass $\Delta M$ from
$\Delta\omega$, the position of the particle must be known.  Proposed
position determination schemes~\cite{Boisen_2005, Hansen_2007,
  Craighead_2008, Craighead_2009} rely on detectors to measure the
frequency shifts of several vibration modes. While this poses no
problems in principle, it causes practical difficulties for NEM-MS
operating in the GHz regime.

We propose a detection scheme that only requires measurements in a
single narrow band centered at the fundamental mode resonance
frequency of a square 2D resonator. Our method uses the nonlinear
response of the resonator by exploiting the interaction between
vibration modes to make information about higher modes available at
the fundamental frequency. We illustrate by showing, analytically and
numerically, how the nonlinear response of micrometer-size graphene
resonators~\cite{Atalaya_2008, Hone_2009} can be used for single
particle mass measurements with zeptogram precision at room
temperatures.
\begin{figure}
\epsfig{file=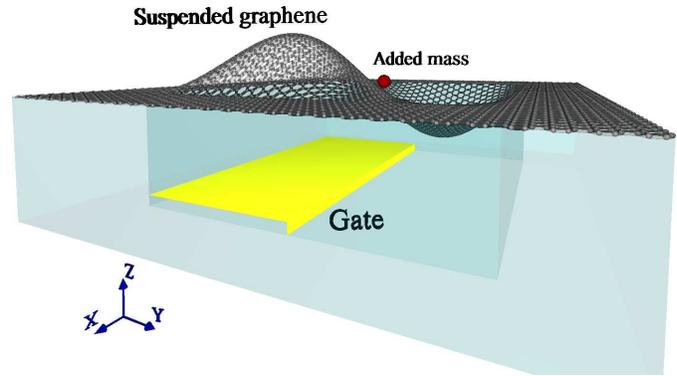, width=\linewidth}
\caption{Possible realisation of a NEM mass spectrometer using a suspended square graphene sheet
with all edges clamped. Below the graphene an electrostatic gate for actuation and transduction
is placed symmetrically with respect to the X-axis and asymmetrically with respect to the Y -
axis. By electrostatic actuation of vibration modes, a mass $\Delta$M located at an arbitrary position
${\bf X}_M = (X_M, Y_M)$ can be determined.
\label{fig:fig0}} \vspace*{-0.3cm}
\end{figure}

Several other technology tracks are being considered for NEM-MS
devices. One is downscaling of Si-MEMS~\cite{Datskos_2003,
  Roukes_2004_1, Roukes_2006, Roukes_2007, Roukes_2009} where the
present state-of-the-art give a minimum detectable mass of $\sim 10$
zg~\cite{Roukes_2009}. Another track relies on carbon nanotubes
(CNTs)~\cite{deHeer_1999} and has already reached sub-zg
levels~\cite{Zettl_2006, Bachtold_2008_1, Bockrath_2008, Zettl_2008}.
However, after the discovery of graphene~\cite{Novoselov_2004}, novel
2D NEMS devices have been explored~\cite{Mceuen_2007,Bachtold_2008_2,
  Houston_2008, McEuen_2008}, including mass detectors with zg
sensitivity~\cite{Hone_2009}. Apart from increasing the adsorbtion
cross-section, 2D-NEMS can also have degenerate flexural modes. As we
show, this degeneracy makes possible to distinguish single-particle
from multi-particle adsorption. Graphene also represents the ultimate
material for 2D-NEMS through its combination of large strength and low
mass.

\section{System}
We consider a square graphene sheet with mass $M$ and side length
$L_0$ suspended in the $XY$-plane above an actuation gate (See Figure~\ref{fig:fig0}). 
{The sheet is simply clamped at all edges}. The gate
geometry, which has a symmetry line parallel to the Y-axis, is chosen
such that the fundamental and higher order modes can be excited. The
transverse deflection $w({\bf X},t)$ of the membrane is given
by~\cite{Atalaya_2008}
\begin{equation}\label{eq:Duff}
  \rho \ddot{w} +c\dot{w}- \sum_{\xi = X,Y} \partial_{\xi}(T_{\xi} \partial_{\xi}w) = P_z({\bf X},t).
\end{equation}
Here $P_z$ is the external pressure on the sheet. {This
  pressure comes from the electric biasing on the gate electrode. The
  exact geometry of the gate, and the exact ${\bf X}$-dependence of
$P_z$ need not be known. It suffices that $P_z$ has the proper
symmetry.}  And, $T_X=T_Y=T_0 + T_1|\nabla w|^2$ are sheet tension
components where $T_0$ is an initial tension and $T_1 \approx
112$~N/m.  Equation (\ref{eq:Duff}) is nonlinear due to
stretching-induced tension~\cite{Atalaya_2008}.  For a particle with
relative mass $\epsilon \equiv \Delta M/M$ adsorbed at ${\bf X}_M$,
the density is $\rho({\bf X})=\rho_0+\Delta M\delta({\bf X}-{\bf
  X}_M)$, where $\delta({\bf X})$ is the 2D delta function and
$\rho_0$ is the density of graphene.

{For future convenience, we begin by rescaling Eq.(\ref{eq:Duff}) into a dimensionless
form. We do this by} introducing the length and
time scales $h_0=L_0\sqrt{T_0/T_1}$ and $t_0=L_0\sqrt{\rho_0/T_0}$, we
write the deflection as {$u({\bf x},\tau)=w(L_0{\bf x}, t_0 \tau)/h_0$.}
{Equation~(\ref{eq:Duff}) then becomes
\begin{equation}\label{eq:Duff2}
 [1+\epsilon\delta({\bf x}-{\bf x}_M)]\ddot{u} +\gamma\dot{u}- \nabla^2 u-\sum_{\xi = x,y} \partial_{\xi}(|\nabla u|^2 \partial_{\xi}u) = p_z
\end{equation} 
where $\gamma=ct_0/\rho_0$ and}  $p_z=P_zt_0^2/(\rho_0h_0)$.

\section{Linear response}
\begin{figure}[t]
\epsfig{file=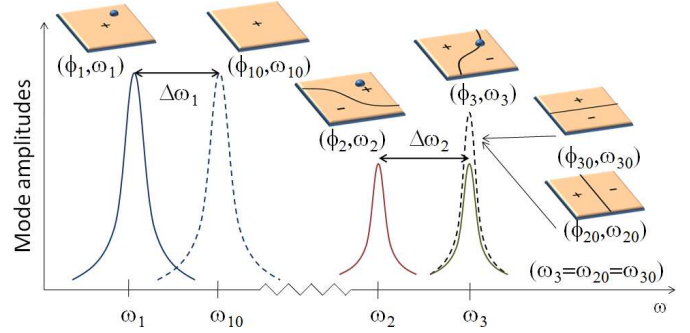, width=\linewidth}
\caption{Amplitudes for the three lowest flexural
  eigenmodes as functions of drive frequency $\omega$ for weak
  driving. {\bf Dashed lines:} Linear response without added mass.
  The unperturbed mode shapes $\phi_{10}$, $\phi_{20}$ and $\phi_{30}$
  are indicated on the plaquettes where the locations of nodelines
  antinodes are shown. The modes $\phi_{20}$ and $\phi_{30}$ are
  degenerate.  {\bf Solid lines:} Linear response in the presence of
  an added mass.  The mode functions are $\phi_1$, $\phi_2$ and
  $\phi_3$ with shapes indicated on the plaquettes. The blue dots show
  the position of the added mass.
\label{fig:fig1}} \vspace*{-0.3cm}
\end{figure}

We consider first small deflections where $T_{X,Y}\approx T_0$, and
the resonator is in the linear regime. 
The eigenmodes are then determined from
\begin{equation}
-\omega^2[1 + \epsilon \delta({\bf x}-{\bf x}_M)]u-\nabla^2 u=0,\quad {\bf
  x}\in[0,1]^2.
\label{eq:eig}
\end{equation}
Without adsorbed particles $\epsilon=0$, the first three mode shapes are
$\phi_{10} = 2\sin\left({\pi x}\right)\sin\left({\pi y}\right)$,
$\phi_{20} = 2\sin\left({2\pi x}\right)\sin\left({\pi y}\right)$,
$\phi_{30} = 2\sin\left({\pi x}\right)\sin\left(2{\pi y}\right)$,
with eigenfrequencies $\omega_{10}^2=2\pi^2$ and $\omega_{20}^2=\omega_{30}^2=5\pi^2$.
To linear order in $\epsilon$, adding a mass at ${\bf x}_M$ leads to
$\omega_1^2=\omega_{10}^2(1-\epsilon \bar{\phi}_1^2)$,
$\omega_2^2=\omega_{20}^2(1-\epsilon{\cal N}^2)$, and
$\omega_3=\omega_{30}$.  Here $\bar{\phi}_m\equiv \phi_{m0}({\bf
  x}_M)$ and {${\cal N}\equiv[\bar{\phi}_{2}^2 +
\bar{\phi}_{3}^2]^{1/2}$}. To zeroth order in $\epsilon$, $\phi_1=\phi_{10}$,
$\phi_2=[\bar{\phi}_{2}\phi_{20}+\bar{\phi}_{3}\phi_{30}]/{\cal N}$
and $\phi_3=[{\bar{\phi}_{2}\phi_{30}-\bar{\phi}_{3}\phi_{20}}]/{\cal
  N}$. These solutions are illustrated in Fig.~\ref{fig:fig1}.

For a two-fold degenerate mode, the frequency of one mode is lowered
due to particle adsorbtion. The other mode will not change frequency
since it has a nodal line passing through the location $\textbf{x}_M$.
This allows a simple test to see if more than one particle has been
adsorbed. A multi-particle adsorption results in frequency shifts for
both the initially degenerate modes.

\section{Nonlinear response}
To study the nonlinear dynamics of the system, we expand the
{scaled} deflection {$u$ in Eq.(\ref{eq:Duff})} in the eigenmodes $\phi_m({\bf x})$ of the linear problem
[Eq.(\ref{eq:eig}) with $\epsilon\neq 0$] as $u({\bf
  x},\tau)=\sum_{m=1}^\infty u_m(\tau)\phi_m({\bf x})$. 
This yields a
system of coupled Duffing equations for the mode amplitudes $u_m$
\begin{equation}\label{eq:ode}
D_{m}(\ddot {u}_m+\omega_m^2 u_m)+\gamma\dot{u}_m+ \sum_{rst=1}^{\infty} A_{mrst} u_r u_s u_t =p_m.
\end{equation}
Here $D_m=1+\epsilon\phi_m({\bf x}_M)^2=1+\epsilon\tilde\phi_m^2$,  $A_{mrst}=\int d{\bf x}\,(\nabla \phi_m \cdot \nabla
\phi_r)(\nabla \phi_s \cdot \nabla \phi_t)$,
and $p_m=\int d{\bf x}\,\phi_m p_z$. 
{As $\epsilon\ll 1$ we have to lowest order in $\epsilon$, $D_m^{-1}\approx 1-\epsilon\tilde\phi_m^2\approx \omega_m^2/\omega_{m0}^2$. 
\begin{eqnarray}\label{eq:ode2}
&&(\ddot {u}_m+\omega_m^2 u_m)+\gamma[1-\epsilon\tilde\phi_m^2]\dot{u}_m\nonumber\\
&&+\sum_{rst=1}^{\infty} A_{mrst}[1-\epsilon\tilde\phi_m^2] u_r u_s u_t =p_m[1-\epsilon\tilde\phi_m^2].
\end{eqnarray}}

{In what follows we will } consider the weakly nonlinear
regime.  The cubic nonlinearities in Eq.~(\ref{eq:ode2}) can be then
be treated {using the method of averaging (Krylov-Bogoliubov
  method)}.  {In this method, both the damping
  $\gamma\dot{u}$, the driving $p_m$, and the terms of order $u^3$ are
  of the same order and small (see for instance Ref.~\cite{Nayfeh}).
  Formally, $\gamma$ can in this method be treated as a small
  parameter of a perturbation expansion.  To simplify the analysis,
  terms of order $\cal{O}(\epsilon\gamma)$ can then be considered as
  higher order terms and omitted.}  Further, only drive frequencies
close to $\omega_{10}$ and $\omega_{20}=\omega_{30}$ are used and
equations for the three lowest modes suffice.  These approximations
give
\begin{eqnarray}
&&\ddot {u}_1+\gamma\dot{u}_1+(\omega_1^2+5[A u_2^2+A u_3^2]) u_1+Au_1^3=p_1\nonumber \\
&&\ddot {u}_2+\gamma\dot{u}_2+(\omega_2^2+5[Au_1^2+Cu_3^2])u_2+Bu_2^3=p_2\nonumber \\
&&\ddot {u}_3+\gamma\dot{u}_3+(\omega_3^2+5[Au_1^2+Cu_2^2]) u_3+Bu_3^3=p_3\label{eq:123}
\end{eqnarray}
where $A=5\pi^4$,
$B={161\pi^4}/4+{3\pi^4}\bar{\phi}_2^2\bar{\phi}_3^2/(2{\cal N}^4)$
and $C\approx 41\pi^4/5$. {The ultimate justification for the
  approximations leading up to Eq.~(\ref{eq:123}) are the comparisons
  of the theoretical treatment of the system~(\ref{eq:123}) with the
  numerical simulations of the full equations~(\ref{eq:ode})}.

For the external force of the form {$p_z({\bf
  x},\tau)=p(\tau)g({\bf x})$ where $g$ obeys the symmetry relation
$g({\bf x})=g(|x-0.5|,y)$, the source terms can be written as
\begin{eqnarray}
&&p_1(\tau)=D_1p(\tau)\nonumber\\
&&p_2(\tau)=D_2p(\tau)\cos(\pi y_M)\nonumber\\
&&p_3(\tau)=D_2p(\tau)\cos(\pi x_M)\nonumber.
\end{eqnarray}
Here 
$$D_{1}=2\int d{\bf x}, \sin(\pi x)\sin(\pi y)g({\bf x})$$
and 
$$D_{2}=2\frac{\int d{\bf x}\,\sin(\pi x)\sin(2\pi y)g({\bf x})}{\sqrt{\cos^2\pi
    x_M+\cos^2\pi y_M}}.$$}

{In the expressions for for the source terms $p_n$, the form
  of the driving force, $g({\bf x})$ is included in the coefficients
  $D_{1,2}$. We again stress that the exact form of $g({\bf x})$ is
  not important, and need not be known, as long as it has the symmetry
  property $g({\bf x})=g(|x-0.5|,y)$. It is this symmetry property
  which causes the same coeffecient $D_2$ to appear in both the source
  terms $p_2$ and $p_3$. Hence, any measurable quantity which depends
  only on the ratio $p_2/p_1$ will thus be a function of only the
  particle position ${\bf x_M}$. This will be used in the mass
  measurment scheme presented below.}
\begin{figure}
\epsfig{file=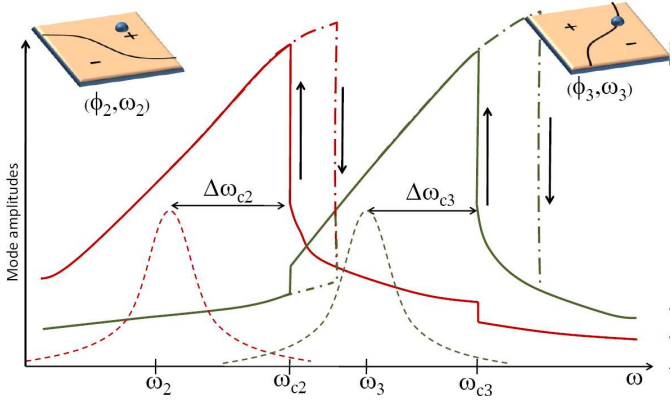, width=\linewidth}
\caption{
  Amplitudes for modes 2 and 3 as functions of drive frequency
  $\omega$ for a square membrane with an added mass.  {\bf Solid
    lines:} Nonlinear response.  {\bf Dashed lines:} Linear response
  (see Fig.~\ref{fig:fig1}). By driving both modes into the nonlinear
  regime, the parameter $r$ [see Eq.~\ref{eq:res2}] can be obtained
  from the frequency shifts $\Delta\omega_{c2}$ and
  $\Delta\omega_{c3}$. The parameter $r$ defines the nodal line of
  mode 3.  Both $\omega_{c2}$ and $\omega_{c3}$ are measured by
  sweeping $\omega$ downwards.  Solid curves were obtained by
  numerical integration of Eq.~(\ref{eq:ode}) with a mass fraction
  $\Delta M/M=0.08$\% located at $(x_M,y_M)=(0.81,0.20)$ (quality
  factor $Q_1=3000$). {\bf Dash-dotted line:} Above the frequencies
  $\omega_{c2,c3}$ hysteretic behavior can be observed by sweeping
  $\omega$ upwards.
  \label{fig:fig3}}  \vspace*{-0.3cm}
\end{figure}

\section{Mass measurement}
To determinine the position of the adsorbed mass we will use the parameters
$r$ and $s$ defined as
\begin{eqnarray}
r &\equiv& \cos(\pi y_M)^2/\cos(\pi x_M)^2\\
s &\equiv& 1 - [\cos^2\left({\pi x_M}\right)+\cos^2\left({\pi y_M}\right)].
\end{eqnarray}
The quantity $s$ is related to the frequency shifts in the linear response regime through
\begin{equation}
1-s\approx\frac{1}{10}\frac{\omega_{20}^2-\omega_2^2}{\omega_{10}^2-\omega_1^2}.
\label{eq:res1}
\end{equation}
{This parameter can thus be determined by applying a weak
  harmonic drive of the form $p(\tau)=\cos(\omega\tau)$ and monitoring
  the location of resonances. Driving the system harder, still with a
  single frequency, puts it in the non-linear regime. However, for a
  single frequency excitation in the weakly non-linear regime, the
  coupling between the equations in (\ref{eq:123}) can be ignored and
  the system turns into three uncoupled Duffing
  equations.
\begin{eqnarray}
&&\ddot {u}_1+\gamma\dot{u}_1+\omega_1^2 u_1+Au_1^3=p_1\nonumber \\
&&\ddot {u}_2+\gamma\dot{u}_2+\omega_2^2 u_2+Bu_2^3=p_2\nonumber \\
&&\ddot {u}_3+\gamma\dot{u}_3+\omega_3^2 u_3+Bu_3^3=p_3\label{eq:123s}
\end{eqnarray}
Characteristic for a driven Duffing oscillator in the nonlinear regime
is the bistability region in parameter space where the system
oscillates with either small or large amplitude depending on the
initial conditions. This leads to the characteristic hysteresis loops
seen in figure~\ref{fig:fig3}.}

The parameter $r$ can be related to the frequency shifts by noting
that the ratio of the forces $p_2(\tau)$ and $p_3(\tau)$ in Eq.
(\ref{eq:123}) is given by $\sqrt{r}$. {As shown in appendix,} the edges of the
hysteresis loops depend on the applied forces as
$(\omega_{cn}^2-\omega_n^2)^3 \approx (9/4)^2 B p_n^2$ ($n$ = 2, 3) so
that
\begin{equation}
r=\left(\frac{\omega_{c2}^2-\omega_{2}^2}{\omega_{c3}^2-\omega_{3}^2}\right)^3.
\label{eq:res2}
\end{equation}
Hence, frequency measurements in the linear and nonlinear regimes can
be used to determine $r$ and $s$. From $r$ and $s$ the position of the
adsorbed particle can be deduced (up to symmetry of the structure).
Knowing the position (in terms of $r$ and $s$) allows calculation of
the mass responsivity ${\cal R}_1$ of the fundamental mode $\phi_1$
{by calculating the linear frequency shift}
\begin{equation}{\cal R}_1({\bf
  x}_M)\approx-2{\omega_{10}}\frac{(s+r)(1+rs)}{(1+r)^2}
\label{eq:res3}
\end{equation}
which gives the added mass $\Delta M=\epsilon M={\cal R}_1^{-1} M
\Delta\omega_1$.
%Figure~\ref{fig:fig4} shows a single-particle mass sensing scheme
%based on Eqs.~(\ref{eq:res1},\ref{eq:res2},\ref{eq:res3}).

{The result presented here rests on three main equations
  (\ref{eq:res1}), (\ref{eq:res2}) and (\ref{eq:res3}). To obtain this
  result we have made two crucial assumptions relating to the symmetry
  of the system; the symmetry leading to mode degeneracy and the
  symmetry of the gate. In any real situation, these symmetries will
  not be exact and it is relevant to question to what extent these
  symmetries will need to be fulfilled. For a complete error-analysis,
  one must analyze the detailed reasons for lifting the degeneracies.
  While such a detailed analysis is beyond the scope of the present work, some observations can be readily made. 
  Firstly, the most crucial symmetry is that of the membrane. For
  the scheme presented here to be relevant thus puts constraints on the
  intrinsic mode splitting $\Delta\omega_{23}\equiv
  \omega_{30}-\omega_{20}$.  The first of these constraints is
  $\Delta\omega_{23}\ll\omega_{20}-\omega_2.$ When this inequality is
  fulfilled, the effect of an adsorbed particle on the nearly
  degeneraty modes is larger than the effect of imperfections leading
  to the intrinsic splitting.  A second criterion, which is less
  obvious, is that
  $$\Delta\omega_{23}\ll \omega_{3}-\omega_{2}$$
  This criterion means that mode 3 does not shift appreciably when the particle is added.}

\section{Narrowband scheme}
Above, we have demonstrated that frequency measurements can be used to
determine the position and mass of the adsorbed particle. We now show
that, by exploiting the nonlinearities in the system, this information
can be obtained by measuring only in a narrow frequency band near the
fundamental mode frequency $\omega_{1}$.

\begin{figure}
\epsfig{file=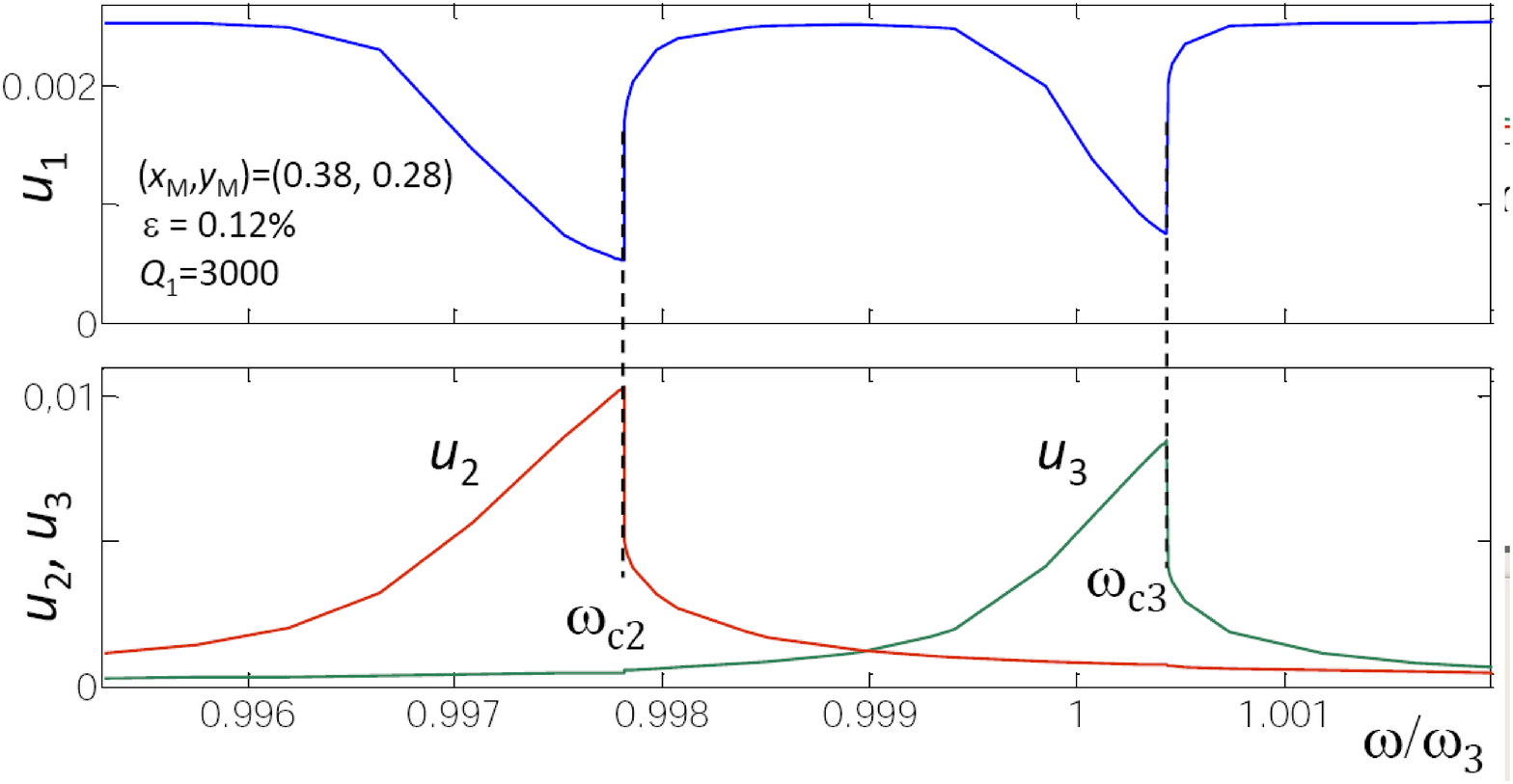, width=\linewidth}
\caption{
  Mode amplitudes obtained by numerical integration of the
  system~(\ref{eq:ode}) using a gate signal $p_A\cos(\omega_1 t)+p_B\cos(\omega t)$.  {\bf Upper panel:} Amplitude
  of mode 1 as function of variable drive frequency $\omega$. {\bf
  Lower panel:} Amplitudes for modes 2 and 3 as functions of drive
  frequency~$\omega$. The frequency $\omega_1$ is fixed at the resonance of mode
  1 while $\omega$ is varied. Due to nonlinearity the modes couple.
  This causes the resonant frequency of mode 1 to depend on the
  amplitudes of modes 2 and 3. It will thus shift away from $\omega_1$ for
  finite amplitudes of modes 2 and 3.  Hence, by measuring the response
  of mode 1, the responses of modes 2 and 3 can be probed by measuring
  only in a narrow frequency band around $\omega_1$.
  \label{fig:fig2}}  \vspace*{-0.5cm}
\end{figure}

Equations~(\ref{eq:123}) represent a system of
three coupled Duffing oscillators for the modes amplitudes $u_n$
[$n=1,2,3$]. Here, the effective resonant frequency of a mode depends
not only on the oscillation amplitude of the mode itself but also on
the amplitudes of other modes so that for instance $\omega_1^2$
increase by approximately $5A \sum_{2,3} \langle u_k^2 \rangle$ where
$\langle \cdot \rangle$ denotes time-average over an oscillation
period. This allows us to choose to use the fundamental mode to
monitor the amplitudes of modes 2 and 3 as follows: In the first step,
the system is excited with a single frequency signal $p(\tau) = p_A
\cos(\omega \tau)$ and the frequency $\omega_1$ of the fundamental
mode in the linear regime is determined. The frequency of this
excitation, and detection, is henceforth kept fixed at $\omega_1$. A
second excitation signal $p_B\cos(\omega \tau)$ is superimposed on the
signal at frequency $\omega_1$. When the amplitude $p_B$ is low, the
excitation of mode 2 in the linear regime for $\omega = \omega_2$ can
be detected as a reduction of the oscillation amplitude of the
fundamental mode. This is because the effective frequency of the
fundamental mode is shifted away from $\omega_1$ due to the excitation
of mode 2. Finally, when $p_B$ is increased, the mode 2 is driven into
the nonlinear regime and $\omega_{c2}$ can be determined. Similarly,
$\omega_3$ and $\omega_{c3}$ can be obtained. The effect of the mode
interaction between the fundamental mode and modes 2 and 3 are shown
in Fig.~\ref{fig:fig2}.

At first hand one may object to this scheme by noting that when the
fundamental mode is strongly excited, it affects the frequencies
$\omega_2$ and $\omega_{c2}$. However, since both $\omega_2^2$ and
$\omega_{c2}^2$ shift by the same amount, these shifts cancel out (to
first order) in the expression for $r$. The cancellation occurs also
in the expression for $s$ if the resonant frequencies $\omega_{n0}$
before mass adsorption are determined through the same narrowband
scheme.

%\begin{figure}[t]
%\epsfig{file=MS.eps, width=\linewidth,clip}
%\caption{Single-particle mass sensing (MS) scheme. The resonance frequency shift,
% $\Delta \omega_1$, of the fundamental mode is used to determine whether analytes are adsorbed on
%the resonator. The resonance frequency shift of mode 3, $\Delta \omega_3$, is used to
%determine whether a single analyte is adsorbed on the analyte. A second particle outside the
%nodal line created by the first particle would lead to a shift $\Delta \omega_3 \neq 0$. If
%a single analyte is on the membrane, we probe the nonlinear response of modes 2 and 3 to determine
% the parameters $r$ and $s$. The latter parameters determine the position, $\textbf{x}_M$, of the added particle
% and also the resposivity ${\cal R}_1$ of mode 1. Then, the value of the added mass $\Delta M$ follows from
% $\Delta M=\epsilon M={\cal R}_1^{-1} M \Delta\omega_1$. Note that all quantities needed in this
% MS scheme can be measured at a narrow frequency band centered at the fundamental mode using
% the mode interaction effect as shown in Fig.~\ref{fig:fig2}. It is also shown the schematics
% of a graphene-based nanoelectromechanical device for mass sensing.
% \label{fig:fig4}}
%\end{figure}
\begin{figure}[t]
\epsfig{file=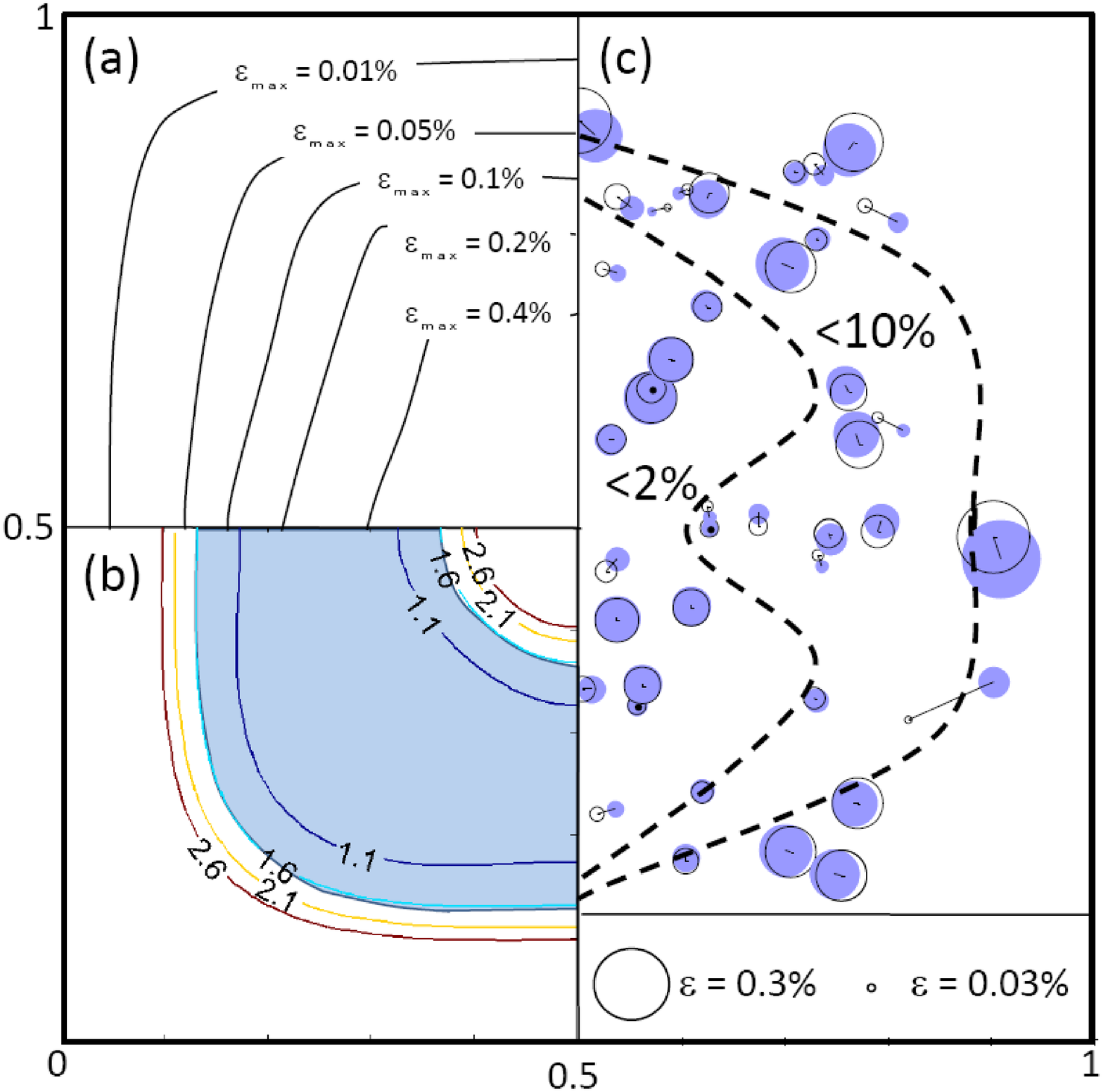, width=\linewidth}
\caption{
  {\bf (a)} Maximal values of $\epsilon\equiv \Delta M/M$ due to
  limitations of first order perturbation theory.  Within each
  contour, mass fractions up to $\epsilon_{\rm max}$ can be determined with a
  5\% accuracy. {\bf (b)} Contours of minimum $\epsilon Q_1$ where
  Eq.~(\ref{eq:res2}) is applicable. E.g., in the shaded area
  Eq.~(\ref{eq:res2}) is valid for $\epsilon>1.6/Q_1$.  {\bf (c)}
  Determination of randomly deposited masses using numerical
  integration of Eq.~(\ref{eq:ode}) for a membrane with $Q_1=3000$.
  The masses were uniformly distributed in the range $0.02\%
  <\epsilon< 0.35 \%$. Frequencies were determined using an accuracy
  of $|\Delta\omega/\omega|\approx 0.5\cdot10^{-4}$. The positions of
  the deposited masses are shown by shaded symbols. The open symbols
  were obtained using Eqs.~(\ref{eq:res1}) and
  (\ref{eq:res2}).  The size of the markers are proportional to
  $\epsilon$. The dashed lines indicate regions where $|(\epsilon-\epsilon_{\rm exact)}/{\epsilon_{\rm exact}}|$ is
  less than 2\% or 10\%.
  \label{fig:fig5}}  \vspace*{-0.3cm}
\end{figure}

\section{Mesurement sensitivity and range}
We now consider {\em sensitivity} and {\em range}. In the NEM-MS experiments reported in the literature~\cite{
Datskos_2003,Roukes_2004_1, Roukes_2006, Roukes_2007, Roukes_2009, deHeer_1999, Zettl_2006,Bachtold_2008_1, Bockrath_2008, Zettl_2008}, the sensitivity is usually
taken as the smallest detectable mass. In our case this occurs when the particle is adsorbed at the sweet spot of the resonator at ${\bf
  x}_M=(0.5, 0.5)$.  This leads to
$\Delta M_{\rm min}=0.5(\Delta\omega_1/\omega_{10})_{\rm min}M$.  The
intrinsic limitation on $|\Delta \omega/\omega|$ comes from
thermomechanical noise that determines how small resonance shift can be reliably detected. If the detector bandwidth $\Delta \omega$ is narrower than the resonance at
$\omega_1$ we have $|\Delta\omega/\omega|>Q_1^{-1} 10^{-{\rm DR_n}/20}$~\cite{Roukes_2004_1}. Here ${\rm DR}_n$ is the dynamic range
of mode $n$ and $Q_1$ the quality factor of the fundamental mode. For
modes $n=1,2,3$ we find
$${\rm
DR}_n=10\log_{10}\left[\frac{R_n}{Q_1}\left(\frac{T_0}{T_1}\right)\frac{T_0
    L_0^2}{k_B T}\right],$$
where $R_1\approx 0.6$ and $R_2=R_3\approx
0.3$. For a device with $L_0=1\,\mu$m,
$Q_1=3000$ and $\omega_{1}/(2\pi)= 2$ GHz we find $\Delta M_{\rm min}\approx
\frac{1}{2}M Q_1^{-1} 10^{-2.5}\approx 0.5$ zg at $T=300$ K. At lower temperatures the sensitivity improves as $T^{1/2}$.

Thermal fluctuations also influence the determination of the
frequencies $\omega_{c2,c3}$. If the system performs low-amplitude
oscillations with $\omega$ close to $\omega_c$, thermal fluctuations
can cause transitions to the high-amplitude state before $\omega_c$ is
reached.  To accurately determine $\omega_c$ we must have $W\ll
\omega_c$ where $W$ is the rate for transitions to the high amplitude
state.  This rate obeys $W\propto e^{-R E_T/(k_BT)}$
where $E_T\equiv {T_0^2 L_0^2}/T_1$ \cite{Dykman_1994}. As demonstrated in
Ref.~\cite{Cleland_2005_2}, the strong exponential dependence of
$W$ on system parameters can for NEMS lead to an enhanced sensitivity in
the measurements of $\omega_{c2,c3}$ compared to the frequency
measurements in the linear regimes.

We now consider the {\em range} of masses that can be reliably {\em measured} with the nonlinear
mass determination scheme presented above. This must not be confused with the
{\em sensitivity} discussed above which only considers the {\em minimum detectable} mass change.
The range includes both upper and lower
bounds on $\epsilon\equiv\Delta M/M$.
The upper bound arises from omitting terms of ${\cal O}(\epsilon^2)$ and higher in the relation
$\Delta\omega_1={\cal R}_1\epsilon+{\cal O}(\epsilon^2)$.
Fig.~\ref{fig:fig5}a shows contours on a quadrant of the unit square corresponding to the membrane.
Each contour encloses a region where the relative error due to
omitting terms of ${\cal O}(\epsilon^2)$ is less than 5\%.  For
instance, masses with $\epsilon$ up to $\epsilon_{\rm max}=0.1\%$ can
only be determined with a relative error less than 5\% if they are
located inside the $\epsilon_{\rm max}=0.1\%$-contour.  The upper
bound can be improved upon by using numerically calculated values of
$\Delta\omega_1(\epsilon,{\bf x}_M)$ instead of perturbation theory.

Specific to this scheme is that to determine $r$ in
Eq.~(\ref{eq:res2}), the regions of multivalued response for modes 2
and 3 must not overlap. Not only will an overlap lead to frequency
shifts (the jump in amplitude of mode 3 at $\omega=\omega_{c2}$ in
Fig.~\ref{fig:fig3} comes from such a shift), but we have also
observed that it leads to richer dynamics, including Hopf bifurcations
with limit cycles~\cite{Lauterborn_1995}. The necessary criterion for
non-overlap can be shown [using Eq.~(\ref{eq:123})] to give
a lower bound $\epsilon_{\rm min} \gtrsim
2.2[{\cal N}({\bf x}_M)]^{-2}Q_1^{-1}$. Fig.~5b shows contours of constant values of
$\epsilon_{\rm min} Q_1$. There, regions close to the edges
and the center are excluded. Because the responsivity ${\cal
  R}_1(r,s)\rightarrow 2\omega_{10} s+{\cal O}([1-s]^2)$ as
$s\rightarrow 1$, the exclusion of the central area is superficial.
For example, if we want to use the part of the membrane with $0.1<x,y<0.9$,
we have approximately the lower bound $\epsilon \gtrsim 3 Q^{-1}$.
For a square membrane of 1~$\mu$m side ($M \approx 760$~ag), the present scheme
is applicable to masses larger than $\Delta M_{\textrm{min}} \approx 0.76$~ag
(assuming $Q=3000$).

\section{Numerical simulations}
To test the scheme we implemented an automated mass measurment
algorithm which numerically integrated the system~(\ref{eq:ode}) with
a randomly deposited mass on the membrane.  The algorithm then
determined the frequencies $\omega_{1,2,3}$ and $\omega_{c2,c3}$ and
calculated $\epsilon$ using Eqs.~(\ref{eq:res1}),(\ref{eq:res2}), and
(\ref{eq:res3}). The results are shown in Fig.~\ref{fig:fig5}c.  The
relative error in $\epsilon$ ranges from 0.1\% to 98\% with the larger
errors near the edges where $\epsilon$ is highly sensitive to
position. Masses close to the edges could be identified by overlapping
responses for modes 2 and 3 in the nonlinear regimes and were
discarded.  As can be seen, the errors in position of the remaining
particles are typically small.

\section{Conclusions}
In conclusion, we have proposed a scheme to determine both the
position and mass of a single particle adsorbed on a vibrating
graphene membrane. We have shown that by using bimodal excitation and
exploiting the nonlinear response of the resonator, measurements can
be restricted to a narrow frequency band near the fundamental
frequency. Considering that the typical resonance frequencies of
graphene membranes lie in the GHz range, this simplification offers
significant experimental advantages. {These measurements provide information about the  
resonance frequencies and the coefficients of the nonlinear terms of the dynamic 
equations (Kerr constants) of the high-order modes. In a resonator without 
special symmetries, the mass and position of the adsorbed particle can be
determined using the resonance frequency shifts of three different
modes ---measured at a narrow frequency band near the fundamental frequency.
If the resonator is square, it is possible to separate the single-particle 
adsorbtion events by watching out for changes of the resonance frequency of the third mode.
Using a gate with a proper symmetry, it is possible to determine the mass and position of a
adsorbed analyte on the membrane by using the resonance frequency
shifts of modes 1 and 2 and the frequencies of the lower-edge
 bistability regions of modes 2 and 3.}  \\ 
As an example we have studied a square membrane with an area of 1~$\mu$m$^2$, 
eigenfrequency of 2 GHz and quality factor of $Q\approx 3000$. For this membrane the
sensitivity at room temperature (minimum {\em detectable} mass change)
is below 1 zeptogram with a practical operating range in the attogram
region.  This can be compared with, {\em e.g.}, quartz crystal
microbalances that have mass sensitivities in the nanogram range.

\acknowledgments
We acknowledge the Swedish Research Council and
the Swedish Foundation for Strategic Research for the financial support.
{We also wish to thank Referee B at EPL for valuable comments and criticism.}
{
\section{Appendix} We here present, for completeness, a brief derivation of the location of the bifurcation point on the so called backbone curve for the Duffing oscillator. Similar derivations can be found in most books on nonlinear systems (see for instance~\cite{Nayfeh}).

Consider a harmonically driven Duffing oscillator
$\ddot{x}+2\gamma\dot{x}+\omega_0^2x+\kappa x^3=p_0\cos(\omega t)$
and introduce slowly in time varying action-angle variables $r(t)$ and $\phi(t)$ such that
$x=r\sin(\omega t+\phi)$ and $\dot{x}=r\omega\cos(\omega t+\phi)$.
Substituting these expressions into the differential equation and averaging over
the fast oscillations (see for instance~\cite{Nayfeh}) gives the system
\begin{eqnarray}
\dot{r}\omega&=&-\gamma \omega r -\frac{p_0}{2} \sin \phi \nonumber\\
r\omega\dot\phi&=&\frac{\omega_0^2-\omega^2+(3\kappa/4)r^2}{2}r-\frac{p_0}{2}\cos\phi \nonumber
\end{eqnarray}
The frequency response curve is found by solving for the stationary regime $\dot{r}=\dot{\phi}=0$. This amounts to solving the frequency response equation
\begin{equation}\label{eq:app1}
4\gamma^2 r^2\omega^2+r^2\left[(\omega_0^2-\omega^2)+\frac{3}{4}\kappa r^2\right]^2=p_0^2.
\end{equation}
We seek the solution when the bifurcation occur. This is exactly the point where $\frac{\partial \omega}{\partial r}=0$. Using this equality while taking the derivative with respect to $r$ in the frequency response equation (\ref{eq:app1}), leads to an equation for the critical frequency $\omega_c$ (considering here the limit $\gamma\rightarrow 0$) for transition from the low to large amplitude solution
$$\left[(\omega_0^2-\omega_c^2)+\frac{3}{4}\kappa r^2\right]+\frac{3}{2}r^2\kappa=0.$$
Inserting the solution for $r^2$ in Eq.~(\ref{eq:app1}) (still using $\gamma=0$) gives
$$p_0^2=\left(\frac{4}{9}\right)^2\frac{(\omega_c^2-\omega_0^2)^3}{\kappa}.$$}

\end{document}